\documentstyle[12pt,epsf]{article}

\begin{document}

\vspace{0.5cm}

\centerline { \bf  Bose-Einstein condensate in non-homogeneous gravitational field}

\vspace{0.3cm}

\centerline { Igor K. Kulikov  \footnote {electronic address: kulikov@jpl.nasa.gov}} 

\centerline {\small \it Quantum Computing Technologies Group, Jet Propulsion
Laboratory, California Institute of Technology,}  
\centerline {\small \it 4800 Oak
Grove Drive, Pasadena, California 91109-8099 }


\vspace{0.3cm}

\centerline { Abstract}

{\it Ground state properties of trapped Bose condensate 
with repulsive interaction  in non-homoge-neous
gravitational field are studied. Spatial  structure of Bose condensate    
and its momentum distributions 
in 3-D anisotropic trap are considered 
by the solution of the modified non-linear Schr\"{o}dinger equation. 
The results are compared with the corresponding properties 
of condensate in a harmonic trap without gravitational field.}\\ 


\vspace{0.3cm}

Bose-Einstein condensation (BEC) is an outstanding example of a case where
the properties of quantum matter can
be studied in laboratory experiments.
Since the theoretical discovery of BEC in the mid-1920s, most studies 
in this area 
were concentrated on the properties of homogeneous condensates with effectively 
infinite spatial extents.
However, experimental success in achieving and observation of
Bose condensation  in trapped dilute atomic gases,  which was done during 
the last five years, 
provided the possibility for study of the properties of an inhomogeneous confined
 condensate \cite {an1}-\cite{dav1}.
A trapped Bose gas exhibits features that are impossible to observe 
in a homogeneous medium.
One of the remarkable features of these experiments is  the observation of
Bose condensation in coordinate as well as in  momentum space.
The localization in space of the condensed fraction of a low temperature Bose gas
allows the detection of  the condensate and to estimate the size of the BEC sample, 
to find its shape and momentum distribution of its atoms \cite{wie1}-\cite{dur1}. 
Moreover, the ability to manipulate such coherent matter  provided additional 
stimuli
for experimental and theoretical study of this phenomenon.
The properties and behavior 
of a BEC in a trap depend on  the presence of various factors. 
One of these factors is the interaction of the trapped atoms with a gravitational 
field.
Gravity has a dominant influence on the evolution of atoms released from the trap.
The dominance of gravity over the thermal effects
has been successfully used in atom cloud gravimeters \cite{kas1}-\cite{pet1}.
The use of BEC as a gravity and  gravity gradient sensor opens wide 
perspectives for the creation of highly sensitive gravimeters and gravity gradiometers \cite{sna}.  
In this paper we present quantitative results on the influence of a non-homogeneous gravitational
field  on the behavior of a Bose condensate with 
repulsive interactions in a 3-D anisotropic trap. 
We  determine the properties of  a BEC in a gravitational field by studying 
its geometry, momentum distribution, and dynamical properties.  
We concentrate on the results obtained by the 
analysis of
the solutions to the  non-linear Schr\"{o}dinger equation  for trapped atoms in 
external
gravitational field.
Zero temperature  solutions for the condensate are found using the  Gross-Pitaevski (GP) 
equation
\cite{ed1}-\cite{dal1}.
Our objective is to provide a general framework for the analysis of the behavior 
of a confined Bose condensate which interacts with a gravitational field through the 
 modified GP-equation and   to investigate its properties.

A BEC created in a trap is under the influence of different forces, and the properties of
the condensate depend on the combination of three potentials. A magnetic trap generates
a time-averaged harmonic potential, while the second effective potential is formed from the binary
collisions of the atoms. Finally, the condensate is under the influence of 
an external gravitational field which gives rise to a gravitational potential. 
The properties of the condensate at zero temperature are described 
by  GP theory \cite{gin1}-\cite{gp1} which in its standard form includes two potentials. 
Our aim  is to  
modify the GP equation in order to take into account gravitational interactions.\\
The many body Hamiltonian describing N interacting bosons with atomic mass $m$ 
is given by the equation
\begin{eqnarray}
H=\int d {\bf x} \hat\psi^{+}({\bf x})
\left[ -\frac{\hbar^2}{2m}\Delta+V_{ext}({\bf x}) \right]\hat\psi({\bf x})
\nonumber\\
+\frac{1}{2} \int d {\bf x^{'}}d {\bf x}
\hat\psi^{+}({\bf x^{'}})\hat\psi^{+}({\bf x})
V({\bf {x^{'}}-{\bf x}})
\hat\psi({\bf x^{'}}) \hat\psi ({\bf x}), \label{e1}
\end{eqnarray}
where  $\hat\psi ({\bf x})$ and $\hat\psi^{+}({\bf x})$ are the boson field 
operators which annihilate and create a particle at the position ${\bf x}$,
respectively, and $V({\bf {x^{'}}-{\bf x}})$ is two body interatomic potential.
Time evolution for the field operator $\hat\psi ({\bf x},t)$
 is found from the Heisenberg equation 
\begin{eqnarray}
i \hbar \frac{\partial }{\partial t}\hat\psi ({\bf x},t)=[\hat\psi ({\bf x},t),H]
\nonumber\\
=\left[ -\frac{\hbar^2}{2m}\Delta+V_{ext}({\bf x})\right . 
\nonumber\\
\left .  
+ \int d {\bf x^{'}}
\hat\psi^{+}({\bf x^{'}},t)
V({\bf {x^{'}}-{\bf x}})
\hat\psi({\bf x^{'}},t)
\right]\hat\psi({\bf x},t). \label{e2} 
\end{eqnarray}
Only binary collisions at low energy
are relevant for the dilute gas at low temperatures. These collisions are
well described by s-wave scattering processes and allow us
to replace $ V({\bf {x^{'}}-{\bf x}})$ with an effective interaction
\begin{equation}
V({\bf {x^{'}}-{\bf x}})=V_o \delta({\bf {x^{'}}-{\bf x}}), \label{e3}
\end{equation}
where the coupling constant $V_o$ is related to s-wave scattering length
$a$ through
\begin{equation}
V_o=\frac{4 \pi \hbar ^2 a}{m}. \label{e4}
\end{equation}
To overcome the problem of solving the full 
many-body Schr\"{o}dinger equation for interacting systems we can use
the mean field approach \cite{fw1},\cite{noz1}. The mean field approach is based on  
separation
of the condensate fraction of the bosonic field operator. 
For a non-uniform
and time-dependent configuration we have
\begin{equation}
\hat\psi ({\bf x},t)=\phi({\bf x},t)+\hat\psi ^{'}({\bf x},t), \label{e5}
\end{equation}
where $\phi({\bf x},t)$ is a complex function defined as the expectation 
 value for the field operator 
 $\phi({\bf x},t)$=$<\hat\psi ({\bf x},t)>$. 
As a result, the equation for the condensate wave function $\phi({\bf x},t)$ 
is found using (\ref{e2}) and
(\ref{e5}),  and is written in the form 
\begin{equation}
i \hbar \frac{\partial }{\partial t}\phi({\bf x},t)=
\left[ -\frac{\hbar^2}{2m}\Delta+V_{ext}({\bf x})
+V_o |\phi({\bf x},t)|^2 \right]
\phi({\bf x},t). \label{e6}
\end{equation}
This non-linear Schr\"{o}dinger type equation  is called  the Gross-Pitaevski (GP) 
equation. 
The validity of this equation is based on the conditions that the s-wave 
scattering length is smaller than the average distance between atoms, 
and that the  number of atoms in the condensate is large.
To obtain the GP-equation for the ground state,  one can write the condensate wave 
function  as 
\begin{equation}
\phi({\bf x},t)=\phi({\bf x})e^{-i\mu t/\hbar}, \label{e7}
\end{equation}
where the wave function $\phi({\bf x})$ is normalized to the total number of
particles $\int d {\bf x} |\phi({\bf x})|^2=N$, and $\mu$ 
is a chemical potential. Inserting  (\ref{e7}) into (\ref{e6}), we find
the equation for the wave function $\phi({\bf x})$
\begin{equation}
\left[ -\frac{\hbar^2}{2m}\Delta+V_{ext}({\bf x})
+\frac{4\pi \hbar^2 a}{m}|\phi({\bf x})|^2 \right]
\phi({\bf x})=
\mu \phi({\bf x}). \label{e8}
\end{equation}
The external potential for the  confined atoms is the sum of the trap potential
and a contribution due to the interaction with gravitational field: 
$V_{ext}=V_{trap}+V_g$. 
The magnetic trap  generates a harmonic confining potential of the form
\begin{eqnarray}
V_{trap}({\bf x})=
\frac{m}{2}\left( \omega_x^2 x^2+\omega_y^2 y^2+\omega_z^2 z^2 \right ), 
\label{e9}
\end{eqnarray}
where $\omega_i$ are the projections of the angular trap frequencies on 
the coordinate axis, and vector ${\bf x}=(x,y,z)$ defines  the displacement 
from the trap center ${\bf x_o}=(x_o,y_o,z_o)$.
The gravitational interaction is described by $V_{g}({\bf X})=m\Phi({\bf X})$, where 
the potential for the given mass distribution is found as
$\Phi({\bf X})=-G\sum_im_i/|{\bf X}_i-{\bf X}|$. 
 The potential of a gravitational field is an unobservable 
value.  
The intrinsic characteristics that uniquely characterize the gravitational field in 
Newtonian gravity are the first and the second spatial derivatives of the potential. 
They form the vector of gravitational acceleration $g_i=-\partial \Phi/\partial x_i$, and 
the gravity gradient (or E\"{o}tv\"{o}s) tensor
\begin{equation}
\Gamma_{jk}=\frac{\partial ^2 \Phi}{\partial x^j \partial x^k}. \label{e10}
\end{equation}
Each component of this tensor defines the acceleration difference along 
the direction $j$ per unit separation along the direction $k$. The gravity gradient
tensor is symmetric, and its trace is related to the local mass density by Poisson's
equation. These conditions define the number of its independent components.  
In the local coordinate basis of the trap the gravitational potential 
can be written with the help of the vector gravitational acceleration  
and the gravity gradient tensor as 
\begin{equation}
\Phi({\bf x_o},{\bf x})=\Phi({\bf x_o}) -g_j({\bf x_o})x^j
+\frac{1}{2} \Gamma_{jk}({\bf x_o})x^jx^k. \label{e11}
\end{equation}
We will locate the coordinate system of the trap in such a way as  to obtain the components of 
the vector of gravitational acceleration as 
\begin{equation}
g_x=0, ~~ g_y=0, ~~ g_z =-|{\bf g}|, \label{e12}
\end{equation}
In this approximation, the leading components of the gravity gradient tensor are
$\{\Gamma_{xx},\Gamma_{yy},\Gamma_{zz}\}$. The components of the gravity gradient are not 
independent. They are connected with each other 
by the Laplace equation which leads to  
$\Gamma_{xx}+\Gamma_{yy}=-\Gamma_{zz}$. 
In the given coordinate system, the equation for the external potential is
\begin{eqnarray}
V_{ext}({\bf x})=
\frac{m}{2}\left [\left(\omega_x^2+\Gamma_{xx} \right) x^2 + 
\left(\omega_y^2+\Gamma_{yy} \right) y^2 
+\left(\omega_z^2+\Gamma_{zz} \right) z^2 \right]
+mg z+m\Phi_o. \label{e13}
\end{eqnarray}
Consider an axially symmetric confined potential with
 angular frequencies 
$\omega_x=\omega_y=\omega_{\perp}$ 
and $\omega_z=\lambda\omega_\perp$, where $\lambda$ is an asymmetry parameter.
Then, the symmetry of the trap allows us to write down GP equation as
\begin{eqnarray}
\left[ -\frac{\hbar^2}{2m}\Delta+
\frac{m \omega_\perp^2}{2}\left( \gamma_x^2(\Gamma)x^2+\gamma_y^2(\Gamma)y^2
+\gamma_z^2(\Gamma) z^2 \right)   \right . 
\nonumber\\ 
\left . +mgz
+\frac{4\pi \hbar^2 a}{m}|\phi(\bf x)|^2 \right]
\phi({\bf x})=
\left( \mu_g-m\Phi_o\right) \phi({\bf x}), \label{e14}
\end{eqnarray}
where   
\begin{eqnarray}
\gamma_x^2(\Gamma) =\left( 1+\frac{\Gamma_{xx}}{\omega_\perp^2} \right),
\nonumber\\
\gamma_y^2(\Gamma) =\left( 1+\frac{\Gamma_{yy}}{\omega_\perp^2} \right),
\nonumber\\
\gamma_z^2(\Gamma) =\lambda^2\left( 1+\frac{\Gamma_{zz}}{\omega_z^2} \right).
\label{e15}
\end{eqnarray}
In the equation (\ref{e15}), the factors 
$\{\gamma_x(\Gamma),\gamma_y(\Gamma),\gamma_z(\Gamma)\}$, are functions of the diagonal
components of gravity gradient tensor.
 In order to simplify the analysis of the equation 
(\ref{e14}), we will scale
the  coordinates using a spatial scale parameter  
$\zeta=\left( {\hbar}/{m\omega_\perp}\right)^{1/2}:$  
\begin{equation}
{\bf x}_1=\zeta^{-1}{\bf x}. \label{e16}
\end{equation}
We  also introduce the scaled wave function 
\begin{equation}
\phi({\bf x})=\left( N \zeta^{-3}\right)^{\frac{1}{2}}\phi_1({\bf x}_1), \label{e17}
\end{equation}
which satisfies the normalization condition  
\begin{equation}
\int d^3{\bf x}_1 |\phi_1({\bf x}_1)|^2=1. \label{e18}
\end{equation}
In the new variables, the dimensionless GP-equation (\ref{e14}) 
has the form
\begin{eqnarray}
\left[ -\Delta_1+
\gamma_x^2(\Gamma)x_1^2+\gamma_y^2(\Gamma)y_1^2+\gamma_z^2(\Gamma) z_1^2 +2D(g)z_1\right .
\nonumber\\
\left . 
+u_1|\phi_1(x_1,y_1,z_1)|^2 \right]
\phi_1(x_1,y_1,z_1)=2\mu_1\phi_1(x_1,y_1,z_1), \label{e19}
\end{eqnarray}
where 
\begin{equation}
u_1=\frac {8 \pi a N}{\zeta} \label{e20}
\end{equation}
is the dimensionless interaction strength,
\begin{equation}
\mu_1=\frac{1} {\hbar \omega_\perp}(\mu_g-m  \Phi_o) \label{e21}
\end{equation}
is a dimensionless eigenvalue of  the equation, and 
\begin{equation}
D(g)=\frac{mg \zeta}{\hbar \omega_\perp}=g\left( \frac{m}
{\hbar \omega_\perp ^3} \right)^{\frac{1}{2}} \label{e22}
\end{equation}
is a  linear function of gravity.
The dimensionless form of GP equation (\ref{e19}) 
is suitable for integration by means of  appropriate methods.
 The non-linear term in this equation 
does not allow us to obtain analytical solutions, and   numerical
computer simulation has to be used instead.  However, it is possible 
to find analytical solutions in two limiting cases: 
non-interacting particles and strongly repulsive interactions.\\
{ \it (a) Condensate without interaction}\\  
If we neglect two-body collisions,  solving the  GP equation 
can be reduced to finding the solution of (\ref{e19}) with $u_1=0$ 
\begin{eqnarray}
\left[ -\Delta_1+
\gamma_x^2(\Gamma)x_1^2+\gamma_y^2(\Gamma)y_1^2
+\gamma_z^2(\Gamma)\left( z_1+\frac{D(g)}{\gamma_z^2(\Gamma)} \right)^2
 \right]
\phi_1(x_1,y_1,z_1)
\nonumber\\
=2\tilde{\mu_1}\phi_1(x_1,y_1,z_1), \label{e23}
\end{eqnarray}
where 
\begin{equation}
\tilde{\mu}_1(g,\Gamma)=\mu_1+\frac{1}{2}\left(\frac{D(g)}
{\gamma_z(\Gamma)}\right)^2. \label{e24}
\end{equation}
The dimensionless ground state solution is
\begin{eqnarray}
\phi_1(x_1,y_1,z_1) =A \exp \left[-\frac{\gamma_x(\Gamma)x_1^2}{2} -
\frac{\gamma_y(\Gamma)y_1^2}{2}\right]\times
\nonumber\\
\exp \left[-
\frac{\gamma_z(\Gamma)}{2}\left(z_1+\frac{D(g)}{\gamma_z^2(\Gamma)}\right)^2 \right], \label{e25}
\end{eqnarray}
where the normalization constant is found from condition (\ref{e18})
as $A=(\gamma_x\gamma_y\gamma_z/\pi^3)^{1/4}$, and
\begin{equation}
\tilde\mu_1 (g,\Gamma)=\left(\gamma_x(\Gamma)+ \gamma_y(\Gamma)
+\gamma_z(\Gamma)\right)/2. \label{e26}
\end{equation}
The ratio of the mean values of momenta can be found using the  wave
function
\begin{eqnarray}
\phi_1(p_x,p_y,p_z) =\exp\left[ -\frac{\zeta^2}{2\hbar^2} \left(
\gamma_x^{-1}(\Gamma)p_x^2+ \gamma_y^{-1}(\Gamma)p_y^2+\gamma_z^{-1}(\Gamma)p_z^2  
\right) \right].  \label{e27}
\end{eqnarray}
From this it follows that the ratios of the mean momenta are
\begin{eqnarray}
\sqrt{\frac{<p_z^2>}{<p_x^2>}}=\sqrt{\frac{\gamma_z(\Gamma)}{\gamma_x(\Gamma)}},~~~
\sqrt{\frac{<p_z^2>}{<p_y^2>}}=\sqrt{\frac{\gamma_z(\Gamma)}{\gamma_y(\Gamma)}},
\nonumber\\
\sqrt{\frac{<p_x^2>}{<p_y^2>}}=\sqrt{\frac{\gamma_x(\Gamma)}{\gamma_y(\Gamma)}}. \label{e28}
\end{eqnarray}
As consequence, the anisotropy of the velocity distribution 
depends on the gravity gradient.
As the Bose condensate is characterized by the absence of thermal excitations, 
 equations (\ref{e28}) reflect the result of atomic motion 
in the trapping potential only.
The transverse and vertical widths of the Gaussian distribution (\ref{e25}) are 
\begin{eqnarray}
<x_1^2>=\frac{1}{2\gamma_x(\Gamma)},~~~<y_1^2>=\frac{1}{2\gamma_y(\Gamma)},~~~ 
<z_1^2>=\frac{1}{2\gamma_z(\Gamma)},  \label{e29}
\end{eqnarray}
and  depend on the gravity gradient.
The average  $<z_1>=-D(g)/\gamma_z^2(\Gamma)$ defines the shift of the condensate 
cloud along the $z$-axis.\\
{\it (b) Condensate in a repulsive limit}\\
The parameter $u_1$ of equation (\ref{e19}) is proportional to the total number
 of  atoms in  the condensate. In the limit of large $N$ one can neglect 
the kinetic energy
term  (the so-called Thomas-Fermi approximation) and rewrite this equation as  
\begin{eqnarray}
\left[\gamma_x^2(\Gamma) x_1^2+\gamma_y^2(\Gamma) y_1^2+
\gamma_z^2(\Gamma) z_1^2 +2D(g)z_1\right .
\nonumber\\
\left . +u_1|\phi_1(x_1,y_,z_1)|^2 \right]
\phi_1(x_1,y_,z_1)=2\mu_1\phi_1(x_1,y_1,z_1). \label{e30}
\end{eqnarray}
The ground state solution of equation (\ref{e30}) is
\begin{equation}
|\phi_1(x_1,y_1,z_1)|^2=
\frac{1}{u_1}\left[2\tilde \mu_1-\gamma_x^2(\Gamma)x_1^2
-\gamma_y^2(\Gamma) y_1^2-\gamma_z^2(\Gamma)\left(z_1
+\frac{D(g)}{\gamma_z^2(\Gamma)} \right)^2 \right], \label{e31}
\end{equation}
where $\tilde \mu_1$ is found from the normalization condition (\ref{e18}) as
\begin{eqnarray}
2\tilde \mu_1(g,\Gamma)=\left[ \frac{15}{8\pi}
\left (\gamma_x(\Gamma)\gamma_y(\Gamma)\gamma_z(\Gamma)\right )u_1\right]^{2/5}
\nonumber\\
=\left[ \frac{15aN}{\zeta}\left (\gamma_x(\Gamma)
\gamma_y(\Gamma)\gamma_z(\Gamma) \right ) \right ]^{2/5}.  \label{e32}
\end{eqnarray}
The chemical potential depends on the trapping frequencies entering the potential $V_{trap}$,
the number of particles $N$, and the components of the gravity gradient.
The  right hand side of  equation  (\ref{e31}) defines the area of the condensate location.
The boundary of the condensate cloud  has the form of an ellipsoid 
with  semiaxes 
\begin{eqnarray}
R_x(g,\Gamma)=\frac{R}{\gamma_x(\Gamma)}, ~~~R_y(g,\Gamma)=\frac{R}{\gamma_y(\Gamma)},~~~
R_z(g,\Gamma)=\frac{R}{\gamma_z(\Gamma)},\label{e33}
\end{eqnarray}
where $R=\sqrt{2\tilde \mu_1(g,\Gamma)}$. As follows from equation (\ref{e31}),
the center of the ellipsoid is shifted 
along the $z$-axis by $-D(g)/\gamma_z^2(\Gamma)$.

In the conclussion we will review the obtained results. 
The expression for wave function (\ref{e25}) in the the approximation of  weak interaction between
atoms allows us to find statistical characteristics of the density distribution (such as first and second 
cumulants, moments, mean and variance along each axis).  
As follows from (\ref{e25}) there is a normal distribution along the $z$-axis as well 
as along the $x$ and $y$-axis. 
The distribution along the $z$-axis is characterized by the moment
$m_1=-D(g)/\gamma_z^2(\Gamma)$, and 
dispersion $\sigma_z^2=(2\gamma_{z}(\Gamma))^{-1}$.
 Therefore, the dispersion $\sigma_z^2$ varies with respect to the $\Gamma_{zz}$ component 
 of the gravity 
gradient tensor, and the density distribution is broadened in the $z$ direction 
due to tidal forces.  
The condensate is shifted in proportion to the gravitational acceleration by
  $<z_1>=m_1=-D(g)/\gamma_z^2(\Gamma)$. 
Similarly, the dispersions of the distributions in $x$ and $y$  
are given by $\sigma_x^2=(2\gamma_{x}(\Gamma))^{-1}$ and 
 $\sigma_y^2=(2\gamma_{y}(\Gamma))^{-1}$.
 This shows that the density distribution is squeezed in the transverse direction 
 due to the gravitational 
 interaction of the atoms with the gravitational field. The central density of this distribution
 is  
 $n(0)=|\phi(0)|^2=(N/\zeta^3)\left[(\gamma_x(\Gamma)
 \gamma_y(\Gamma)\gamma_z(\Gamma))/\pi^3\right ]^{1/2}$ 
 and depends on the 
 number of particles in the condensate $N$ and the components of the gravity gradient through the $\gamma$-factors.
The anisotropy in the velocity distributions 
$\sqrt{<v_z^2>/<v_x^2>}=\sqrt{\gamma_z(\Gamma)/\gamma_x(\Gamma)}$ 
and 
$\sqrt{<v_z^2>/<v_y^2>}=\sqrt{\gamma_z(\Gamma)/\gamma_y(\Gamma)}$   
for non-interacting condensates is given by the ratio of $\gamma$-factors
(Eq. \ref{e28}), and therefore is connected with the variation of the diagonal components 
of the gravity gradient. In spite of the axial symmetry of the trap potential, there is an
asymmetry in the ratio
$\sqrt{<v_y^2>/<v_x^2>}$, which is given by the ratio of $\gamma$-factors 
$\sqrt{\gamma_y(\Gamma)/\gamma_x(\Gamma)}$.  
As follows from  equation (\ref{e31}), the shape of the condensate in the strongly 
repulsive limit also depends on the diagonal components of the gravity gradient 
tensor.   
In the  strongly  repulsive limit, the 3-D shape of the condensate 
is an ellipsoid with semiaxes given by equation (\ref{e33}).
The ellipsoid is squeezed in the transverse  $(x-y)$ plane, and extended 
in the axial $(z)$ direction due to the contributions of the gravity gradient. The ellipsoid 
is also shifted in the $z$ direction due to gravity. 
The energy per particle can be found from the equation $\tilde \mu=\partial E/\partial N$ 
and is given by
$E/N=2/7 \tilde \mu$, where the dimensional chemical potential is $
\tilde \mu=(\hbar \omega_\perp/2)\left[ 15aN
(\gamma_x(\Gamma)\gamma_y(\Gamma)\gamma_z(\Gamma)) /\zeta
\right]^{2/5}$. The contributions to the energy $E$ in this approximation 
are related to the interaction between particles, 
the gravitational interaction and oscillator energies.
The density of particles $n(x,y,z)=|\phi(x,y,z)|^2 $  
is found directly
from equation (\ref {e31}). As follows from equation (\ref{e31}), 
the density distribution vanishes 
in classical turning points 
on the surface of ellipsoid. The central density is   
$n(0)=2\tilde \mu_1 N/(\zeta^3 u_1)$, and depends on the number of particles and the gravity gradient.
The size of the ellipsoid depends on the number of particles $N$ (the semiaxis scale as $N^{1/5}$) and 
the components of the gravity gradient.  
The components of the gravity vector and the gravity gradient 
tensor are defined by the source of the gravitational field. 
For example, for  Earth gravity ($g=9.87~ms^{-2}$
and $\Gamma_{zz}=3\times10^{-6}~s^{-2}$), the correction for $\gamma_z(\Gamma)$ 
depends on the frequency $\omega_z$ (Eq.\ref {e15}) and 
can be numerically estimated 
as $\Gamma_{zz}/\omega_z^2 \sim 10^{-8}$ for $\omega_z \sim 20 Hz$.

Author thanks Chris Adami, Jonathan Dowling, George Hockney, Ulvi Yurtsever
and other members of the JPL Quantum Computing Technologies Group for 
stimulated discussion and comments and gratefully acknowledges 
the support of NRO-DII program.This research was performed at 
the Jet Propulsion Laboratory, California Institute of
Technology, under contract with the National Aeronautic and Space Administration
and was supported by a contract with the National Reconnaissance Office.


\begin{thebibliography}{99}

\bibitem{an1} M.H. Anderson, J.R. Ensher, M.R. Matthews, C.E. Wieman, E.A.
Cornell, "Observation of Bose-Einstein Condensation in a Dilute Atomic Vapor"
Science {\bf 269} (1995) 198-201

\bibitem{br1}C.C. Bradley, C.A. Sackett, J.J. Tollett, R.G. Hulet
"Evidence of Bose-Einstein Condensation in an Atomic Gas with Attractive 
Interactions"
Phys. Rev. Lett. {\bf 75} (1995) 1687-1690 

\bibitem{dav1} K.B. Davis, M.O. Mewes, M.R. Andrews, N.J. van Druten, D.S. Durfee,
D.M. Kurn, W. Ketterle "Bose-Einstein Condensation in a Gas of Sodium Atoms",
Phys. Rev. Lett. {\bf 75} (1995) 3969-3973 

\bibitem{wie1}  C.E. Wieman, D.E. Pritchard, D.J. Wineland "Atom Cooling, Trapping, 
and Quantum Manipulation" Rev. Mod. Phys. {\bf 71} (1999) S253-S262

\bibitem{dur1} D.S. Durfee, W. Ketterle 
"Experimental Studies in Bose-Einstein Condensation"
Optics Express {\bf 2} (1998) 299-313

\bibitem{kas1} M. Kasevich,  S. Chu "Measurement of the Gravitational Acceleration of an Atom 
with a Light Pulse Atom Interferometer" Appl. Phys. B{\bf 54} (1992) 321-332

\bibitem{pet1} A. Peters, K.Y. Chung,  S. Chu 
"Measurement of Gravitational Acceleration by Dropping Atoms"
 Nature {\bf400} (1999) 849-852
 
\bibitem{sna} M.J. Snadden, J.M. McGuirk, P. Bouyer, K.G. Haritos, M.A. Kasevich
"Measurement of the Earth's Gravity Gradient with an Atom
Interferometer-Based Gravity Gradiometer", Phys. Rev. Lett. {\bf 81} (1998) 971-974

\bibitem{ed1} M. Edwards, K. Burnett "Numerical Solution of the Nonlinear Schr\"{o}dinger 
Equation for Small Samples of Trapped Neutral Atoms"
Phys. Rev. A{\bf 51} (1995) 1382-1386

\bibitem{ed2} M. Edwards, R.J. Dodd, C.W. Klark, P.A. Ruprecht, K. Burnett 
"Properties of Bose-Einstein Condensate in an Anisotropic Harmonic Potential"
Phys. Rev. A{\bf 53} (1996) R1950-R1953

\bibitem{bau1} G. Baym, C.J. Pethick "Ground-State Properties 
of Magnetically Trapped Bose-Condensed Rubidium Gas" 
Phys. Rev. Lett. {\bf 76} (1996) 6-9

\bibitem{dal1} F. Dalfovo, S. Stringari "Bosons in Anisotropic Traps: 
Ground State and Vortices" Phys. Rev. A{\bf 53} (1996) 2477-2485 

\bibitem{gin1} V.L. Ginzburg, L. P. Pitaevski
"On the Theory of Superfluidity" 
Zh. Exp. Theor. Phys. {\bf 34} (1958) 1240 

\bibitem{gros1} E.P. Gross "Hydrodynamics of superfluid condensate"
J. Math. Phys. {\bf 4}  (1963) 195

\bibitem{gp1} F. Dalfovo, S. Giorgini, L.P. Pitaevskii, S. Stringari 
"Theory of Bose-Einstein Condensation in Trapped Gases"
Rev. Mod. Phys. {\bf 71} (1999) 463-512


\bibitem{fw1} A.L. Fetter, J.D. Walecka 
\emph {Quantum Theory of Many-Particle Systems}
(McGraw-Hill 1971)

\bibitem{noz1} P. Nozi\'{e}res, D. Pines 
\emph {The Theory of Quantum Liquids Vol.2}
(Addison-Wesley 1990)

\end{thebibliography}
\end{document}